\documentclass[twocolumn,secnumarabic,amssymb,floatfix,aps]{revtex4-2}

\usepackage{graphicx}% Include figure files
\usepackage{dcolumn}% Align table columns on decimal point
\usepackage{bm,amsmath,amsbsy}     % bold math
\usepackage{braket}
\usepackage{appendix}
\usepackage{xcolor}
\usepackage{array}
\usepackage{booktabs}
\usepackage{soul}
\setstcolor{red}
\usepackage{amsthm}
\usepackage{amsfonts}
\usepackage{float}
\usepackage{pstricks}           % e.g. colors can be used in the text
\usepackage{mathrsfs}
\usepackage{hyperref}           % hyerlinks in PDF
\usepackage{hypernat}           % compress sequential citations
\usepackage[latin1]{inputenc}   % don't bother me when using special characters
\usepackage{subfigure}
\usepackage{slashed}    % \slashed{A}~a\!\!\!/
\usepackage{color}
\usepackage{soul}

\usepackage{adjustbox}
\makeatletter
\newcommand{\thickhline}{%
	\noalign {\ifnum 0=`}\fi \hrule height 1pt
	\futurelet \reserved@a \@xhline
}
\newcolumntype{"}{@{\hskip\tabcolsep\vrule width 1pt\hskip\tabcolsep}}
\makeatother

\begin{document}

\title{Strong-field Herman-Kluk propagator method for high-harmonic generation in molecules}

\author{Phi-Hung Tran}
\author{Hao Quan Truong}
\author{R. Esteban Goetz}
\author{Anh-Thu Le}
\affiliation{Department of Physics, University of Connecticut, 196A Auditorium Road, Unit 3046, Storrs, CT 06269}

\date{\today}

\begin{abstract}
We extend our recently developed semiclassical strong-field Herman-Kluk (SFHK) propagator method to calculate high-order harmonic generation (HHG) in diatomic molecules driven by few-cycle intense laser fields. On the example of applications to H$_2$ and N$_2$, we show that our method, based on a combination of the Herman-Kluk propagator and the strong-field approximation, can provide very accurate results for both HHG yield and phase, nearly identical to those from the exact numerical solutions of the time-dependent Schr\"odinger equation. To compare with experimental measurements, averaging over molecular orientations must be performed. Here we demonstrate a distinct and powerful advantage of the SFHK, as its Monte Carlo sampling for the integration over the alignment distribution can be efficiently combined with the integration over the initial momentum distributions of electron wave-packet right after the tunnel exit. Therefore, the total number of trajectories used for the alignment-averaged HHG spectrum does not increase much compared to that for a single fixed alignment. Similar to atomic targets, the main computational task in the SFHK is to solve the classical Hamiltonian equations for the active electron in the combined electron-target ion potential and electron-laser interaction. The motion of the center of each electron wave packet in the continuum, represented by a coherent state, is governed by an independent classical trajectory so that the computation can be parallelized very efficiently.    
\end{abstract}

\maketitle

\section{Introduction}
High-order harmonic generation (HHG) from molecules in intense lasers has been studied extensively both experimentally and theoretically over the past few decades, see, for example, Ref.~\cite{Krausz:rmp09}. Compared to atomic targets, molecules can provide an additional level of control thanks to the recent progress in molecular alignment techniques \cite{Stapelfeldt:rmp03}. With the advance of high-harmonic spectroscopy, molecular structure information can be extracted from HHG measurements, with a temporal resolution down to femtoseconds, see, for example, Refs.~\cite{Worner:nature10,Worner:science11,Worner:science15,He:AdvPhoton23,He:prl24}.     

In principle, methods based on numerical solutions of the time-dependent density functional theory (TDDFT) or time-dependent Schr\"odinger equation (TDSE) can be used for theoretical treatments of HHG from molecules. However, even within the single-active-electron (SAE) approximation, numerical solution of the TDSE is still quite challenging for molecules, especially polyatomic molecules, due to the multi-center nature of the potential and the large computational resources required for proper treatments of electron dynamics in the continuum under intense lasers. Furthermore, to compare with experimental measurements, one generally needs to perform averaging over the molecular orientation distribution, even under some favorable conditions when macroscopic phase-matching calculation is not needed, for example, for very dilute gases. Computationally, such an averaging typically requires hundreds of calculations for molecules with different fixed alignments. 

It is therefore not surprising that many HHG calculations in practice still rely on approximate methods. One of the most popular methods is the strong-field approximation (SFA), developed by Lewenstein {\it et al} \cite{Lewenstein:pra94} and its variations. However, this method (which is also called the Lewenstein model) has been known to suffer serious inaccuracies, which are associated mostly with the neglect of the interaction between electron in the continuum with the target ion. In fact, in the past few decades, there have been various attempts to improve the Lewenstein model by including this interaction in different ways. Such attempts include, for example, the classical-trajectory Monte Carlo (CTMC) method \cite{Botheron:pra09,Soifer:prl10, Higuet:pra11,Abanador:jpb17}, the quantum trajectory Monte Carlo (QTMC) model \cite{Wang:pra21,Wang:OptExp24}. Other related works using the semiclassical approach includes \cite{Hostetter:pra10,Mauger:pra16,Koch:AnnalsPhys21}. However, despite these efforts, the success of these attempts so far has been quite limited.  We also remark that the quantitative rescattering theory (QRS) \cite{Morishita:prl08,Le:pra09,Lin:jpb10,Lin:book2018} has been quite successful, with various applications reported in the literature. This method can be classified as an extension of the three-step model \cite{Corkum:prl93} and the Lewenstein model, but using exact laser-free photo-recombination dipoles.  

Recently we proposed the semiclassical strong-field Herman--Kluk propagator method (SFHK) \cite{tran2024quantum,wcl3-x52t,mcmanus2025delay} for calculating three-dimensional (3D) photoelectron momentum distributions (PMDs) of atoms and molecules in intense laser fields. The method was later extended for HHG process \cite{Tran:HHGatom-2026}. In essence, the SFHK uses a combination of the SFA (to calculate the electron wave-packet just created in the continuum at the tunnel exit) and the semiclassical Herman-Kluk (HK) propagator \cite{Herman1984,Kay1994_1,Grossmann:PhysLett98} (which propagates each newly born wave-packet in the full combined potential, starting right at the tunnel exit). We have demonstrated that the SFHK accurately reproduced PMDs and HHG spectra obtained from the TDSE for various targets \cite{tran2024quantum,wcl3-x52t,mcmanus2025delay,Tran:HHGatom-2026}. 

The purpose of this paper is to extend the SFHK to HHG calculations for molecular targets. On the example of H$_2$ and N$_2$, we will demonstrate that the SFHK can provide very accurate HHG yield and phase, comparable with the exact numerical solutions of the TDSE. We further demonstrate the efficiency of this method for averaging over the molecular orientation distribution. This offers a great advantage, especially for polyatomic targets. 

It is important to note that the HK propagator was used earlier for HHG process in Refs.~\cite{Sand:prl99,Sand:pra00} for a one-dimensional (1D) model atom. However, it was not a true HHG process, as it did not include the ionization step. Instead, a free electron wave packet was initially prepared in the continuum at a large distance from the atom and the harmonics are emitted when this wave packet collides with the target ion under an intense laser. The same model was later used in Refs.~\cite{Zagoya:pra12,Zagoya:njp12}. A similar approach was used in Zagoya {\it et al} \cite{Zagoya:njp14}, however the trajectories are coupled. The use of independent trajectories in the SFHK offers an important advantage. In fact, primitive parallelization can be used to efficiently perform the calculations for millions of trajectories on different computers.  

The rest of this paper is organized as follows. In Sec.~II.1 we briefly describe the key elements of the SFHK method for HHG. In Sec.~II.2 we describe how averaging over the molecular alignment can be done efficiently within the SFHK. We also describe the model potentials used in our methods and give some basic information for numerical solution of the TDSE in Sec.~II.3 and Sec.~II.4, respectively. Our results and detailed analyses for H$_2$ and N$_2$ are presented in Sec.~III. Finally, we finish our paper with a summary. Atomic units are used throughout unless otherwise indicated.

\section{Theoretical Methods}
\subsection{The strong-field Herman-Kluk (SFHK) method for HHG}  

The SFHK method has been proposed and described in Refs.~\cite{tran2024quantum,Tran:HHGatom-2026} for calculating PMDs of atoms and molecules and HHG of atoms in intense laser fields. Here we only describe the essential points together with minor modifications due to molecular targets. More detail about the method can also be found in Refs.~\cite{wcl3-x52t,mcmanus2025delay}. The high accuracy of the SFHK can be attributed to the ability of the semiclassical HK propagator to take the full account of the atomic (or molecular) potential together with the laser interaction once the electron is emitted into the continuum, and is indicative of the adequacy of the SFA for the ionization step. 

For HHG, the main task is to calculate the time-dependent induced dipole, dipole velocity, or dipole acceleration \cite{Burnett:pra92,Le:pra09,BandraukPRA2009}. For atomic targets, we found that the two forms, dipole and dipole velocity, give nearly identical results \cite{Tran:HHGatom-2026}. For molecules, we found it more convenient to calculate dipole velocity $v_z(t) = - \left\langle\ \Psi(t) \right| k_z \left| \Psi(t) \right\rangle$. Here, the laser is assumed to be linearly polarized along $z$-axis and the state vector $|\Psi(t) \rangle$ is assumed to evolve in time in the laser field starting from the initial ground state $|\psi_0\rangle$. We note that within the SFHK, it is more convenient to work in the laboratory frame directly. Therefore, the state vector $\Psi$ is dependent on the molecular orientation in the laboratory frame. At the microscopic level, the HHG yield is proportional to the Fourier transform of the dipole velocity (or induced dipole or dipole acceleration).  

Under the assumptions that excited states play insignificant roles in the ionization and photo-recombination, as well as the depletion and the Stark shift of the ground state are negligible due to sufficiently low laser intensities, the dipole velocity can be approximated as $v_z(t)  \approx - \left\langle\ \psi_0e^{iI_pt} \right| k_z \left| \psi_c(t) \right\rangle + c.c.$ Here, $|\psi_c(t) \rangle$ is the electron wave packet in the continuum. The continuum wavefunction within the SFHK method was derived in Ref.~\cite{Tran:HHGatom-2026}, which is written in the momentum space as
\begin{equation}
	\begin{split}
		\psi_c & ({\bf{p}},t) \propto \sum_s \iint d{\bf{p}}_{t_0} d{\bf{q}}_{t_0} \braket{{\bf{p}}|{\bf{p}}_t,{\bf{q}}_t} C_{{\bf{p}}_{t_0}{\bf{q}}_{t_0}t}e^{iS_\to} \\
		&\times \dfrac{\left\langle\ {\bf{p}}_{t_0} + {\bf{A}}(t_s) -  {\bf{A}}(t_0) \right| {\bf{r}} \cdot {\bf{E}}(t_s) \left| \psi_0 \right\rangle}{\sqrt{{\bf{E}}(t_s)\cdot [{\bf{p}}_{t_0} + {\bf{A}}(t_s) -  {\bf{A}}(t_0)] }}e^{i(S_\downarrow^0 + {\bf{p}}_{t_0} \cdot {\bf{q}}_{t_0})}.
	\end{split}
	\label{SFHK-wf}
\end{equation} 
Here ${\bf{E}}$ and ${\bf{A}}$ are the electric field and vector potential of laser, respectively. $\braket{{\bf{p}}|{\bf{p}}_t,{\bf{q}}_t}$ is the coherent state basis function in the momentum representation at time $t$, with the center at $({\bf{p}}_t,{\bf{q}}_t)$ in the phase space. The width of the coherent state is fixed at $\gamma=1$ in this work. The evolution of each trajectory, indicated by $({\bf{p}}_t,{\bf{q}}_t)$, is governed by classical Hamilton's equations of motion.  $t_s = t_0 + it_t$ is the complex-valued solution of the SFA saddle-point equation describing tunneling ionization
\begin{equation}
	\frac{[{\bf{k}} + {\bf{A}}(t_s) -  {\bf{A}}(t_0)]^2}{2}+I_p = 0.
	\label{SP-eq}
\end{equation}

Here, $t_0$ is the ionization time, while $t_t$ is the ``tunneling'' time, which gives the ionization probability for the electron to be born in the continuum at time $t_0$ with momentum ${\bf{k}} \approx {\bf{p}}_{t_0}$. The phase $S_\downarrow^0$ in Eq.~(\ref{SFHK-wf}) is the action during tunneling and it is given by $S_\downarrow^0({\bf{k}},t_s) = I_pt_s-\int_{t_s}^{t_0}\frac{[{\bf{k}} + {\bf{A}}(t'') -  {\bf{A}}(t_0)]^2}{2} dt''$. The phase associated with the electron propagation in the continuum is $S_\to({\bf{p}}_{t_0},{\bf{q}}_{t_0},t) = \int_{t_0}^t ({\bf{p}}_{t'} \cdot {\bf{\dot q}}_{t'} - H) dt'$. Note that the pre-exponent factor $C_{{\bf{p}}_{t_0}{\bf{q}}_{t_0}t}$ is fully included as its explicit form can be found in Ref.~\cite{Tran:HHGatom-2026}. Following Refs.~\cite{Yan:prl2010,Lai2015,Brennecke2020} we impose the condition ${\bf{q}}_{t_0} = {\rm Re}[\int_{t_s}^{t_0}{\bf{A}}(\tau) d\tau]$, which gives the position of the tunnel exit. Intuitively, we associate each electron trajectory that is born into continuum at time $t_0$ with momentum ${\bf{k}} \approx {\bf{p}}_{t_0}$ at position ${\bf{q}}_{t_0}$ with a coherent state centered at $({\bf{p}}_{t_0},{\bf{q}}_{t_0})$ in the phase space. Each of this coherent state evolves in time. The second line in Eq.~(\ref{SFHK-wf}) describes the ionization step, while the first line describes the ionized wave-packet propagation in the continuum.

It is therefore convenient to sample the initial phase space points by Monte Carlo algorithm according to $\left|\dfrac{\left\langle\ {\bf{p}}_{t_0} + {\bf{A}}(t_s) -  {\bf{A}}(t_0) \right| {\bf{r}} \cdot {\bf{E}}(t_s) \left| \psi_0 \right\rangle}{\sqrt{{\bf{E}}(t_s)\cdot [{\bf{p}}_{t_0} + {\bf{A}}(t_s) -  {\bf{A}}(t_0)] }}e^{i(S_\downarrow^0 + {\bf{p}}_{t_0} \cdot {\bf{q}}_{t_0})}\right|$.

Taking advantage of the Gaussian integral, the wavefunction in Eq.~(\ref{SFHK-wf}) can be written analytically if we expand the initial wavefunction $\psi_0$ in the basis set of the Gaussian-type orbitals (GTOs) \cite{Hehre:jcp69,Ditchfield:jcp71}, as commonly done in standard quantum chemistry software such as {\em Gaussian} and {\em Gamess} \cite{g09,GAMESS}. In practice, once $\psi_0$ is obtained from numerical solution of the TISE, it is then expanded in a GTOs basis set, in which the expansion coefficients are found by fitting to the numerical $\psi_0$. In this work, we choose 6-311G and cc-pV5Z as the GTOs basis sets for H$_2$ and N$_2$, respectively. The same approach was taken for atomic targets in Ref.~\cite{Tran:HHGatom-2026}.  

HHG yield is calculated using the Fourier transform of dipole velocity, since it is proportional to $|v_z(\omega)|^2\omega^2 \approx |d_z(\omega)|^2\omega^4$. Note that we typically use the Hanning window function $W_H(t) = 0.5\left[1+\cos \left(\frac{\pi t}{\tau_H}\right)\right]$ \cite{Camp:jpb18} in the Fourier transform of dipole velocity 
\begin{equation}
	v_z(\omega) = \int v_z(t) W_H(t) e^{i\omega t} dt.
\end{equation}

\subsection{Monte Carlo integration over Euler angles within the SFHK}

The induced dipole velocity constructed by the SFHK can be written as
\begin{equation}
	\begin{split}
		v_z(t) \propto \sum_s & \iint d{\bf{p}}_{t_0} d{\bf{q}}_{t_0} \tilde{C} \left\langle \psi_0 \right| k_z \left| {\bf{p}}_t,{\bf{q}}_t \right\rangle \\
		& \times \left\langle W \right| {\bf{r}} \cdot {\bf{E}}(t_s) \left| \psi_0 \right\rangle + c.c.,
	\end{split}
	\label{vz_HK}
\end{equation}
where we have introduced short-hand notations $\tilde{C} = C_{{\bf{p}}_{t_0}{\bf{q}}_{t_0}t} \exp{[i(I_pt+S_\to)]}$ and $\left\langle W \right| = \dfrac{\left\langle {\bf{p}}_{t_0}+{\bf{A}}(t_s)-{\bf{A}}(t_0) \right| \exp{[i(S_\downarrow^0+ {\bf{p}}_{t_0}\cdot{\bf{q}}_{t_0})]}}{\sqrt{{\bf{E}}(t_s) \cdot [{\bf{p}}_{t_0}+{\bf{A}}(t_s)-{\bf{A}}(t_0)]}}$ to reduce the equation to a simpler form. Note that we work in the laboratory frame such that the laser polarization is along the $z$-axis. Therefore, molecular orientation dependence is implicit for in Eq.~\ref{vz_HK}. For instance, $|{\bf{p}}_t,{\bf{q}}_t\rangle$ is the coherent state at time $t$, with the center at $({\bf{p}}_t,{\bf{q}}_t)$ in the phase space. The evolution of each trajectory (i.e., the center of the coherent state), is governed by classical Hamilton's equations of motion. Therefore, 
$|{\bf{p}}_t,{\bf{q}}_t\rangle$ depends on the molecular orientation implicitly -- see also the next subsection for the model potential and its dependence on the molecular orientation. 

The angular dependence of the dipole velocity can be conveniently described in terms of Euler angles, $(\alpha,\beta,\gamma)$, which specify the 3D orientation of the molecules in real space. To be more explicit, we therefore rewrite $v_z(t)$ in the following as $v_z(\alpha,\beta,\gamma,t)$.

For a molecular ensemble with a certain alignment distribution $\rho(\alpha,\beta,\gamma)$, the averaged dipole velocity $\bar{v}_z(t)$ can be expressed as an integral of $v_z$ over the Euler angles as
\begin{equation}
	\begin{split}
		\bar{v}_z(t) = \frac{1}{8\pi^2} \iiint d\alpha \sin{\beta} d\beta d\gamma \ \rho(\alpha,\beta,\gamma)v_z(\alpha,\beta,\gamma,t).
	\end{split}
	\label{vz_int}
\end{equation}
Here we use the convention of $\alpha \in [0,2\pi]$, $\beta \in [0,\pi]$, and $\gamma \in [0,2\pi]$. For an isotropic molecular distribution, we have $\rho(\alpha,\beta,\gamma) = 1$. For numerical solution of the TDSE, one needs to calculate the dipole velocity for an individual set of $(\alpha,\beta,\gamma)$, and then evaluate the integral in Eq.~(\ref{vz_int}). For example, if each of the Euler angles is sampled by ten points, then the integration in Eq.~(\ref{vz_int}) requires $10^3$ TDSE calculations.

We can combine the integral over the Euler angles with the integral over the initial $({\bf{p}}_{t_0},{\bf{q}}_{t_0})$ of the standard SFHK method by explicitly substituting $v_z$ given by Eq.~(\ref{vz_HK}) into Eq.~(\ref{vz_int}) to get 
\begin{equation}
	\begin{split}
		\bar{v}_z(t) \propto \sum_s & \idotsint d\alpha d\beta d\gamma d{\bf{p}}_{t_0} d{\bf{q}}_{t_0} \tilde{C} \left\langle \psi^{\alpha,\beta,\gamma}_0 \left| k_z \right| {\bf{p}}_t,{\bf{q}}_t \right\rangle \\
		& \times \sin{\beta} \left\langle W \left| {\bf{r}} \cdot {\bf{E}}(t_s) \right| \psi^{\alpha,\beta,\gamma}_0 \right\rangle + c.c.
	\end{split}
	\label{vz_Monte}
\end{equation}

Equation (\ref{vz_Monte}) now involves multiple integrals and can be efficiently evaluated using the Monte-Carlo algorithm. In practice, the initial points in the extended space of $({\bf{p}}_{t_0},{\bf{q}}_{t_0},\alpha,\beta,\gamma)$ are distributed according to $\left|\sin{\beta} \left\langle W \left| {\bf{r}} \cdot {\bf{E}}(t_s) \right| \psi^{\alpha,\beta,\gamma}_0 \right\rangle\right|$. As a consequence, if there are $N$ randomly independent trajectories in a SFHK calculation, there will be $N$ different sets of Euler angles, in principle. This significantly improves the convergence as compared to the direct use of the Eq.~(\ref{vz_int}).
\\

\subsection{The single active electron model potential}

We treat a molecular target within the single-active-electron (SAE) approximation. For homonuclear diatomic molecules such as H$_2$ and N$_2$, we express the model potential for the active electron in the form  
\begin{equation}
	\begin{split}
		V({\bf{r}}) = -\sum_{i=1,2}\dfrac{Z_\infty + (Z_0 -Z_\infty)e^{-\alpha |{\bf{r}} - {\bf{R}}_i|}}{|{\bf{r}} - {\bf{R}}_i|}.
	\end{split}
	\label{POT}
\end{equation}
Here, ${\bf{R}}_1 = +{\bf{R}}/2$ and ${\bf{R}}_2 = -{\bf{R}}/2$ are the positions of the two nuclei of the molecule, with $R$ being the internuclear distance, chosen to be the bond length for the molecule in the equilibrium geometry. For our convenience in this paper we will work with the laboratory frame (which is attached to the laser), while the molecular orientation can be specified by a set of Euler angles. The other model potential parameters, $Z_\infty$ and $Z_0$, can be found by enforcing asymptotic condition as well as near nuclear core condition. Namely, in those two limits we require the potential to be Coulombic, such that $V(r\to\infty) = -\dfrac{1}{r}$ and $V({\bf{r}} \to \pm {\bf{R}}/2) = -\dfrac{Z_0}{|{\bf{r}} \mp {\bf{R}}/2|}$, with $Z_0$ being the nuclear charge of H atom (or N atom) for H$_2$ (or N$_2$). The remaining parameter, $\alpha$, is obtained by fitting the eigenvalue of the time-independent Schr\"{o}dinger equation (TISE) using the potential given in Eq.~(\ref{POT}) to the known value for the HOMO. These parameters and the binding energies of the HOMO (as well as the HOMO-1 in case of N$_2$) are given in Table~\ref{Table_POT} and Table~\ref{Table_Energy}.  

In principle, we can always improve the quality of the model potential by using more parameters. For our purpose in this paper, we have limit ourselves with the version given by Eq.~(\ref{POT}). We found that this form give sufficiently accurate description for the HOMO in H$_2$ and both the HOMO and HOMO-1 in N$_2$, see Table~\ref{Table_Energy}.

\begin{table}[!h]
	\caption{Molecular parameters used in the model potential. Atomic units are used.}
	\centering
	\label{Table_POT}
	\renewcommand{\arraystretch}{1.5}
	\setlength{\tabcolsep}{15pt}
	\begin{tabular}{ c c c c c } 
		\thickhline
		Target     & $Z_\infty$ & $Z_0$ & $\alpha$ & $R$ \\
		\hline
		H$_2$      & 0.5   & 1.0   & 1.79     & 1.40 \\
		N$_2$      & 0.5   & 7.0   & 1.67     & 2.07 \\
		\thickhline
	\end{tabular}
\end{table}

\begin{table}[!h]
	\centering
	\caption{Symmetries and binding energies (in eV) of the HOMO and HOMO-1 obtained with the model potential.}
	\label{Table_Energy}
	\renewcommand{\arraystretch}{1.5}
	\setlength{\tabcolsep}{10pt}
	\begin{tabular}{ c c c c c } 
		\thickhline
		& \multicolumn{2}{c}{HOMO}  & \multicolumn{2}{c}{HOMO-1} \\
		\hline
		& Symmetry & Energy & Symmetry & Energy \\
		\hline
		H$_2$      & $1\sigma_g$   & 15.40     &  &  \\
		N$_2$      & $3\sigma_g$   & 15.54     & $1\pi_u$ & 17.18 \\
		\thickhline
	\end{tabular}
\end{table}

\subsection{Numerical solution of the time-dependent Schr\"odinger equation for HHG}

The theoretical method used to solve the time-dependent Schr\"{o}dinger equation (TDSE) within the single-active-electron (SAE) approximation for the calculation of HHG in polyatomic molecules was described earlier in Ref.~\cite{Tran:HHGatom-2026}. The suite of codes is formulated to accommodate different gauge choices and dipole representations. Note that for our convenience, TDSE calculations are performed in the molecular frame. This is in contrast to the SFHK, in which the laboratory frame treatment is more convenient. Once the dipoles are obtained, we then convert them to the laboratory frame. 

Here we only summarize the main points. 

In the electric dipole approximation, the TDSE reads,
\begin{equation}
	\begin{split}
		i\dfrac{\partial}{\partial t} \left| \Psi(t) \right\rangle = [\hat{H}_0 + \hat{V}_L(t) + \hat{W}_C] \left| \Psi(t) \right\rangle,
	\end{split}
	\label{TDSE}
\end{equation}
where $\hat{H}_0=-\dfrac{1}{2}\nabla^2+V({\bf{r}})$ is the field-free Hamiltonian and $V({\bf{r}})$ is described in Eq.~(\ref{POT}). The term $\hat{V}_L = -{\bf{\mu}} \cdot {\bf{E}}(t)$ describes the laser-molecule interaction in the length gauge (LG) with the dipole operator ${\bf{\mu}} = -{\bf{r}}$. $\hat{W}_C = -i\gamma_C(r)$ is a complex absorbing potential (CAP) introduced to avoid reflections off the grid boundaries during the time propagation. $\gamma_C(r)$ is the strength of the absorbing potential described in Refs.~\cite{GreenmanPRA2010,Tran:HHGatom-2026}.

To solve Eq.~(\ref{TDSE}), the state vector is expressed as a linear combination of eigenvectors $\left| \Phi_n \right\rangle$ of the field-free Hamiltonian. We use a fourth-order explicit Runge-Kutta scheme with initial condition $\left| \Phi(t_0) \right\rangle$ corresponding to the ground electronic state of the target system.

Once the state vector $\left| \Psi(t) \right\rangle$ is obtained, the HHG spectrum can be calculated from the Fourier transform of the time-dependent expectation values of the dipole $\left\langle{\bf{D}}(t) \right\rangle = -\left\langle \Psi(t) |{\bf{r}}| \Psi(t) \right\rangle$. Dipole velocity $\left\langle{\bf{v}}(t) \right\rangle = \left\langle \Psi(t) |-i\nabla| \Psi(t) \right\rangle$ or dipole acceleration $\left\langle{\bf{a}}(t) \right\rangle = -\left\langle \Psi(t) \right|\nabla\hat{V}_a({\bf{r}})-{\bf{E}}(t)\left| \Psi(t) \right\rangle$ can also be used. In agreement with Ref.~\cite{Bandrauk:pra09}, we have confirmed that the Fourier transform of all these three dipole forms are related by $|{\bf{D}}(\omega)|^2\omega^4 \approx |{\bf{v}}(\omega)|^2\omega^2 \approx |{\bf{a}}(\omega)|^2$ for sufficiently weak laser intensities resulting in small ionization probabilities. We also used the Hanning window function \cite{Tran:HHGatom-2026} in the Fourier transform.

In this work, we use a 1,200-nm wavelength linearly polarized laser pulse with peak intensity near $1.5\times 10^{14}$ W/cm$^2$, duration of 4 cycles of the sine-squared envelope. Typically, the radial grid is set with $r_{max} = 115$ and the number of radial DVR points is $N_{DVR} = 390$. The CAP strength is set to $\gamma_C = 0.005$ a.u. and $r_C = 100$. The angular part of the wavefunction is expressed through a linear combination of spherical harmonics with up to $L_{max} = 110$ with $m_{max} = 60$. Similarly, we expand the model potential in the basis of spherical harmonics with up to $L^V_{max} = 65$. The summations over eigenvectors of $\hat{H}_0$ are truncated to an energy cutoff of $\epsilon_{N_E} = 12$ a.u., ensuring convergence of the HHG spectrum. Finally, the time step of 0.02 a.u. is used. We have checked that the results are converged with respect to these parameters.

\section{Results and Discussions}

\subsection{HHG from aligned H$_2$ and N$_2$}

Here we present our SFHK results and compare them with the TDSE calculations to benchmark our theory. We emphasize again that, to ensure meaningful comparisons, for each target the same model potential is used for both SFHK and TDSE. In particular, for both methods, the initial molecular orbital wavefunction of the active electron, typically taken as the HOMO, is obtained by numerical solutions of the time-independent Schr\"odinger equation (TISE) with the same model potential. For computational convenience within the SFHK, initial wavefunction is further approximated by a combination of Gaussian-type orbitals (GTO). As discussed earlier, for both targets our initial wavefunctions are very similar to the HOMOs obtained from typical quantum chemistry software such as {\em Gaussian} or {\em Gamess}. Unless specifically indicated, the laser is chosen to be linearly polarized, with the wavelength of 1,200-nm and the intensity of $1.5\times 10^{14}$ W/cm$^2$. The laser pulse duration is of four cycles, with a sine-squared envelope, and the carrier-envelope phase $\phi_{CEP}=0$. These laser parameters give the Keldysh parameter near 0.6 for both targets, which implies that the HHG process occurs in the tunneling regime. 

\begin{figure}[tb!]
	\begin{center}
		\includegraphics[width=0.95\linewidth]{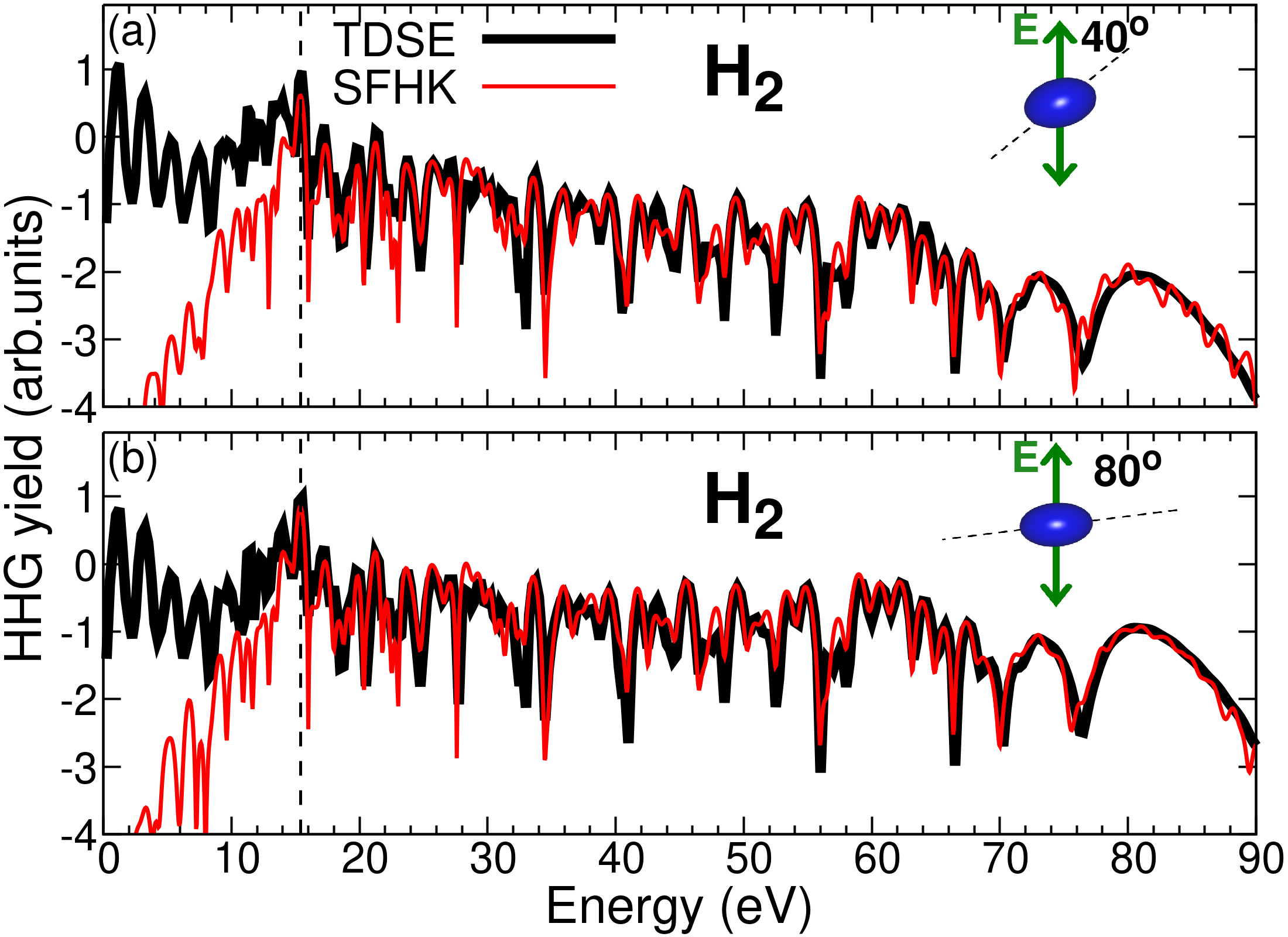}
		\caption{(a) Comparison of HHG spectra from the SFHK (red curve) and the TDSE (black curve) for H$_2$ at the alignment angle $\theta=40^{\circ}$. The vertical black dashed line shows the position of the threshold near 15.4 eV. The HHG yields are shown in the logarithm scale. (b) The same as (a), but for $\theta=80^{\circ}$. For laser parameters, see text.}
		\label{fig:H2-HHG}
	\end{center}
\end{figure}

First, let us start with H$_2$ target. Typical HHG spectra from aligned H$_2$ are shown in Fig.~\ref{fig:H2-HHG} for two alignment angles, $\theta=40^{\circ}$ and $\theta=80^{\circ}$. Clearly, the SFHK results (red line) agree very well with the TDSE (black line), even in fine detail, in the whole range of harmonics energy from the threshold (near 15.4 eV) up to the harmonics cutoff (near 80 eV). Similar level of accuracy for the HHG spectra from the SFHK was also found for all other alignment angles (not shown, but see the discussion of Fig.~\ref{fig:H2-HHG-angular} below). For each alignment angle, we used about $10^{8}$ trajectories in the SFHK calculations to get the converged result. This is almost an order of magnitude more than a typical calculation for an atomic target with similar laser parameters \cite{Tran:HHGatom-2026}. Note that HHG yields below the threshold are not accurate as bound-bound transitions are not included in the current implementation of the SFHK. This is similar to the earlier treatment of the SFHK for atomic targets \cite{Tran:HHGatom-2026}. We further remark that we have used a single overall normalization constant for the SFHK yield to match the TDSE yield for a fixed angle, and that constant factor is used for all other alignment angles.      

To have a better overall quantitative comparison, we show in Fig.~\ref{fig:H2-HHG-angular} HHG yields obtained with the SFHK and the TDSE vs alignment angles for several harmonics, namely, H37, H45, H57, and H77. These harmonics were chosen to represent the lower plateau, middle plateau, upper plateau, and the cutoff harmonics. Clearly, the overall agreement between the SFHK and the TDSE is excellent for HHG yields in the whole range of alignment angle between the molecular axis and the laser polarization direction. Similar comparisons are also shown for the induced dipole phase in the lower panels of Fig.~\ref{fig:H2-HHG-angular}. Overall, general good agreements are seen between the SFHK and the TDSE results. We also found that, in general, convergence of harmonic phases with respect to the number of trajectories used in the SFHK is slower, than for HHG yields. 

\begin{figure}[tb!]
	\begin{center}
		\includegraphics[width=0.95\linewidth]{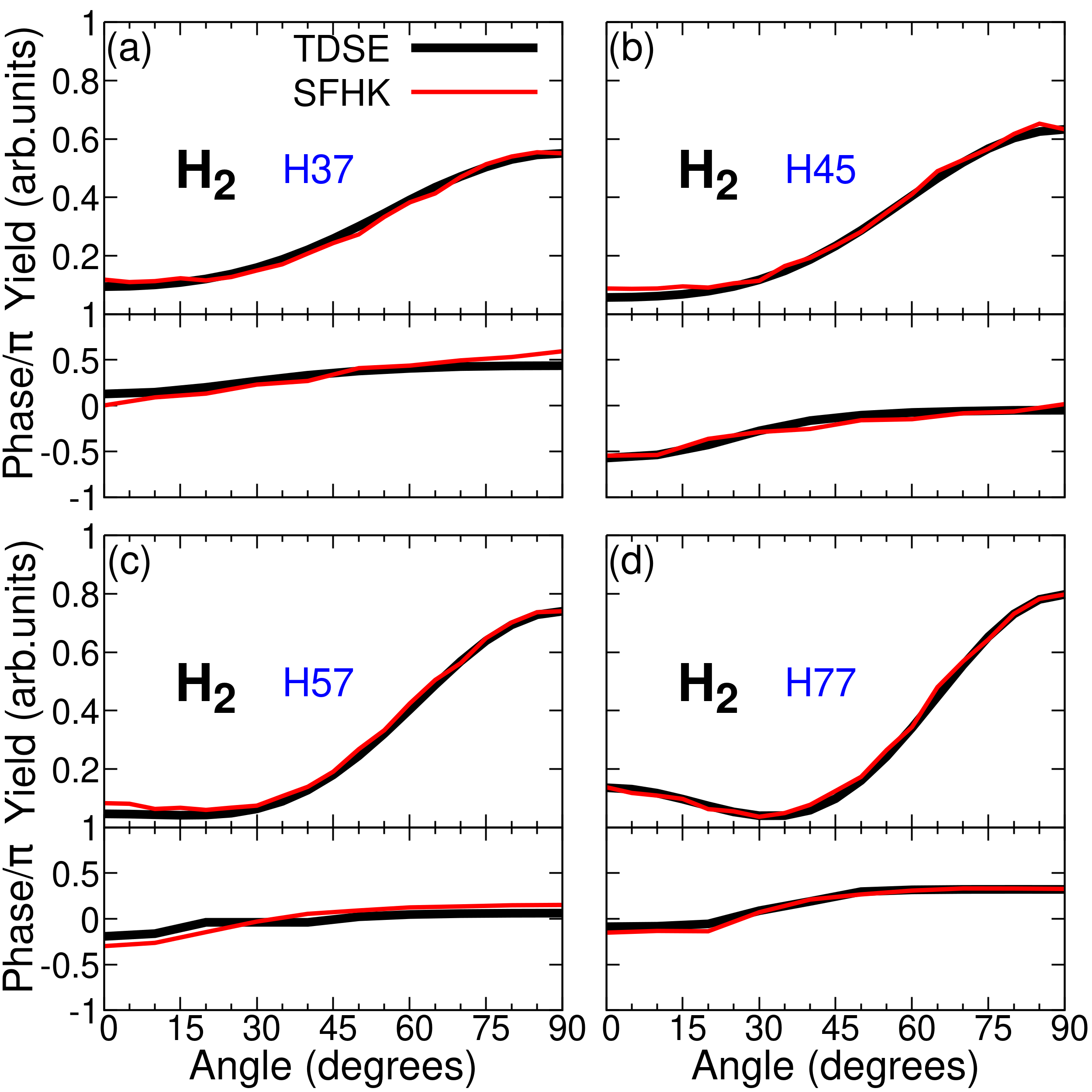}
		\caption{Upper (or lower) panel of (a)-(d): Comparison of HHG yield (or phase) vs alignment angle from the SFHK (red curve) and the TDSE (black curve) for H$_2$ for different harmonics H37, H45, H57, and H77. The laser parameters are the same as in Fig.~\ref{fig:H2-HHG}.}
		\label{fig:H2-HHG-angular}
	\end{center}
\end{figure}

Having established the adequacy of the SFHK for H$_2$, we now move to N$_2$ target. We remark that HHG from aligned N$_2$ has been studied extensively both experimentally and theoretically \cite{Itatani:nature04,McFarland:Science08,Le:jpb09,Jin:pra11,Jin:pra12,Bertrand:prl12,Rupenyan:prl12,Ren:pra13}. It has been shown that for relatively high laser intensities, contributions from multiple molecular orbitals to HHG process are significant, see for example, \cite{McFarland:Science08,Le:jpb09}. In this work we are primarily interested in establishing the validity of the SFHK for molecules. We therefore limit ourselves to the contribution from the highest-occupied molecular orbital (the HOMO), which is of the $\sigma_g$ symmetry.    

\begin{figure}[tb!]
	\begin{center}
		\includegraphics[width=0.95\linewidth]{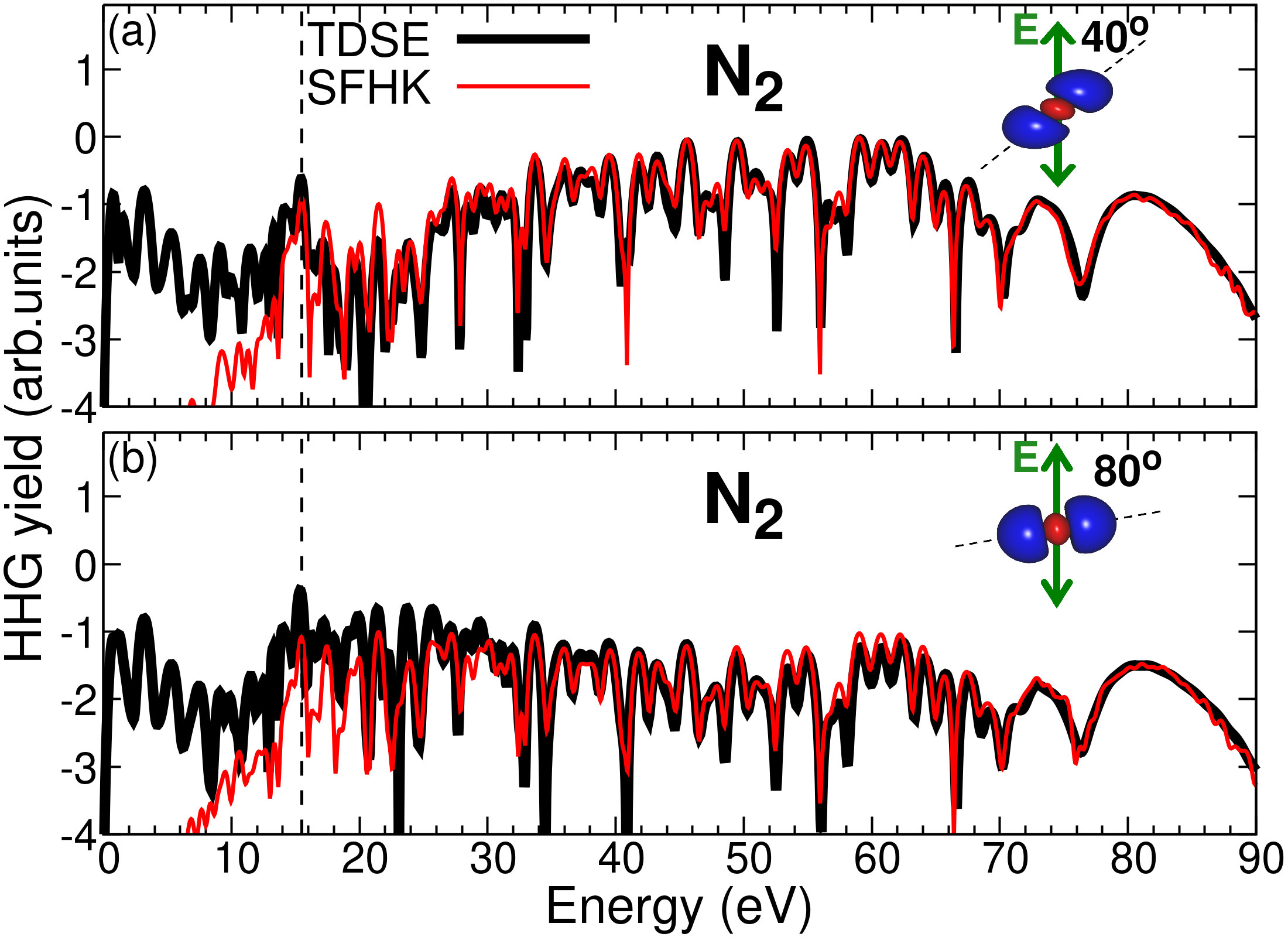}
		\caption{(a) Comparison of HHG spectra from the SFHK (red curve) and the TDSE (black curve) for N$_2$ at the alignment angle $\theta=40^{\circ}$. The vertical black dashed line shows the position of the threshold near 15.5 eV. The HHG yields are shown in the logarithm scale. (b) The same as (a), but for $\theta=80^{\circ}$. The laser parameters are the same as in Fig.~\ref{fig:H2-HHG}.}
		\label{fig:N2-HHG}
	\end{center}
\end{figure}

Typical HHG spectra from aligned N$_2$ are shown in Fig.~\ref{fig:N2-HHG} for two alignment angles, $\theta=40^{\circ}$ and $\theta=80^{\circ}$, under the laser pulse as in Fig.~\ref{fig:H2-HHG}. Again, we see very good agreements between the SFHK and the TDSE results in the whole range of harmonic energy above the threshold at 15.5 eV (indicated by the vertical dashed line). We remark that the similarity in the fine structures of the spectra for those two alignment angles can be interpreted by the QRS as mostly due to the effect of electron-laser interaction. In fact, according to the QRS interpretation, for each alignment angle, the fine structures are superimposed on a generally smooth background, which is proportional to the angle-dependent photo-recombination cross section. This smooth background is largely independent on the laser, but it is strongly dependent on the target.      

\begin{figure}[tb!]
	\begin{center}
		\includegraphics[width=0.95\linewidth]{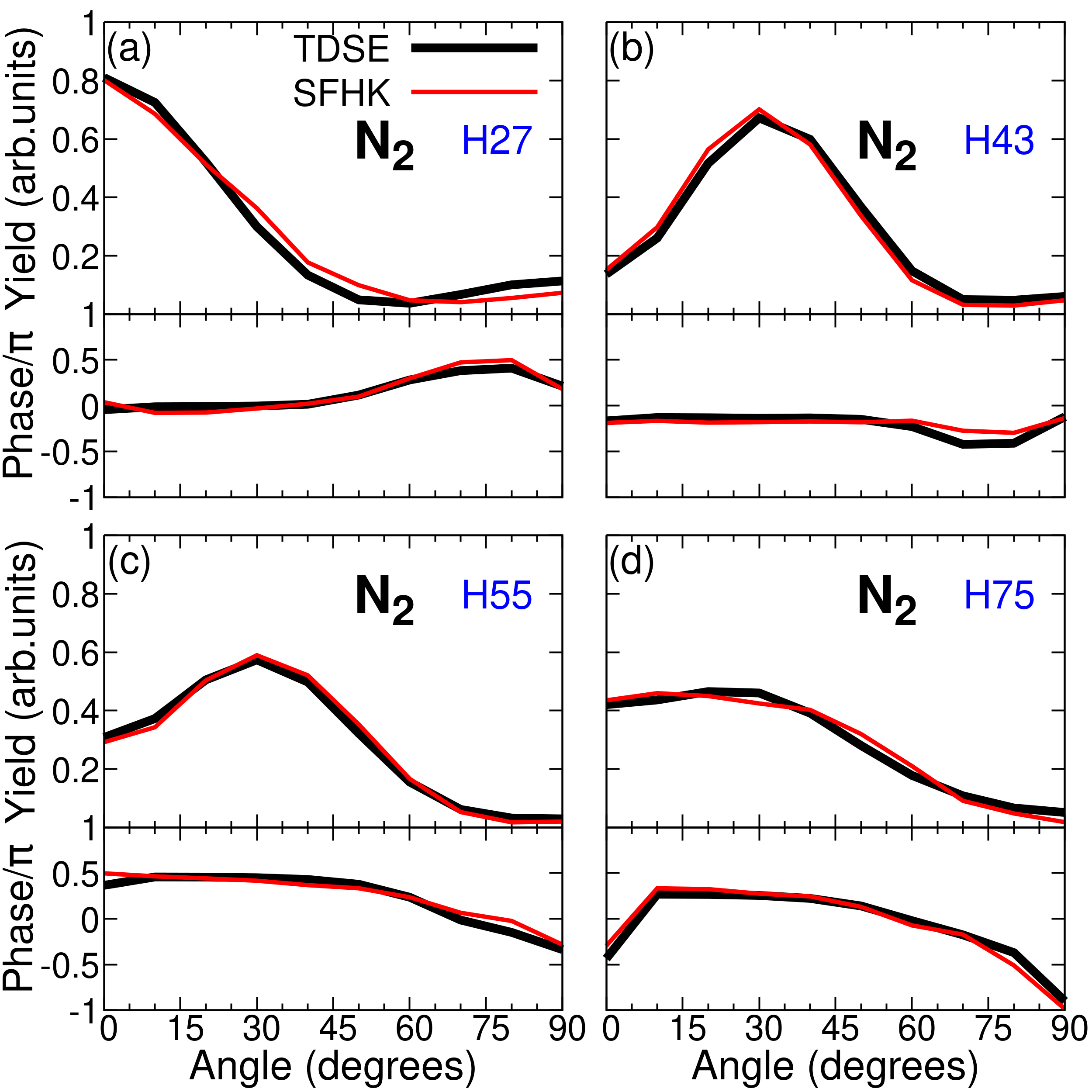}
		\caption{Upper (or lower) panel of (a)-(d): Comparison of HHG yield (or phase) vs alignment angle from the SFHK (red curve) and the TDSE (black curve) for N$_2$ for different harmonics H27, H43, H55, and H75. The laser parameters are the same as in Fig.~\ref{fig:H2-HHG}.}
		\label{fig:N2-HHG-angular}
	\end{center}
\end{figure}

We found a similar good agreement between the SFHK and the TDSE for other alignment angles, except for small angles near $\theta = 0^{\circ}$. In fact, for $\theta < 20^{\circ}$, the HHG spectra from the SFHK are still accurate except for the energy range of $[30:45]$ eV. Interestingly, this is in contrast to the H$_2$ case, in which the SFHK is in a good agreement with the TDSE for any alignment angle. We therefore provide detailed analyses for the case of N$_2$ with $\theta = 0^{\circ}$ in a separate subsection below.  

Angle-dependent yield and phase for some selected harmonics, H27, H43, H55, and H75, are shown in Fig.~\ref{fig:N2-HHG-angular}, in the upper and lower panel for each harmonic, respectively. Again, the agreement between the SFHK and the TDSE is very good. We further remark that the angle-dependent HHG yield and phase reported here are in general agreement with earlier calculations based on the QRS theory \cite{Le:pra09,Le:jpb09,Jin:pra11,Jin:pra12}

\begin{figure}[tb!]
	\begin{center}
		\includegraphics[width=0.95\linewidth]{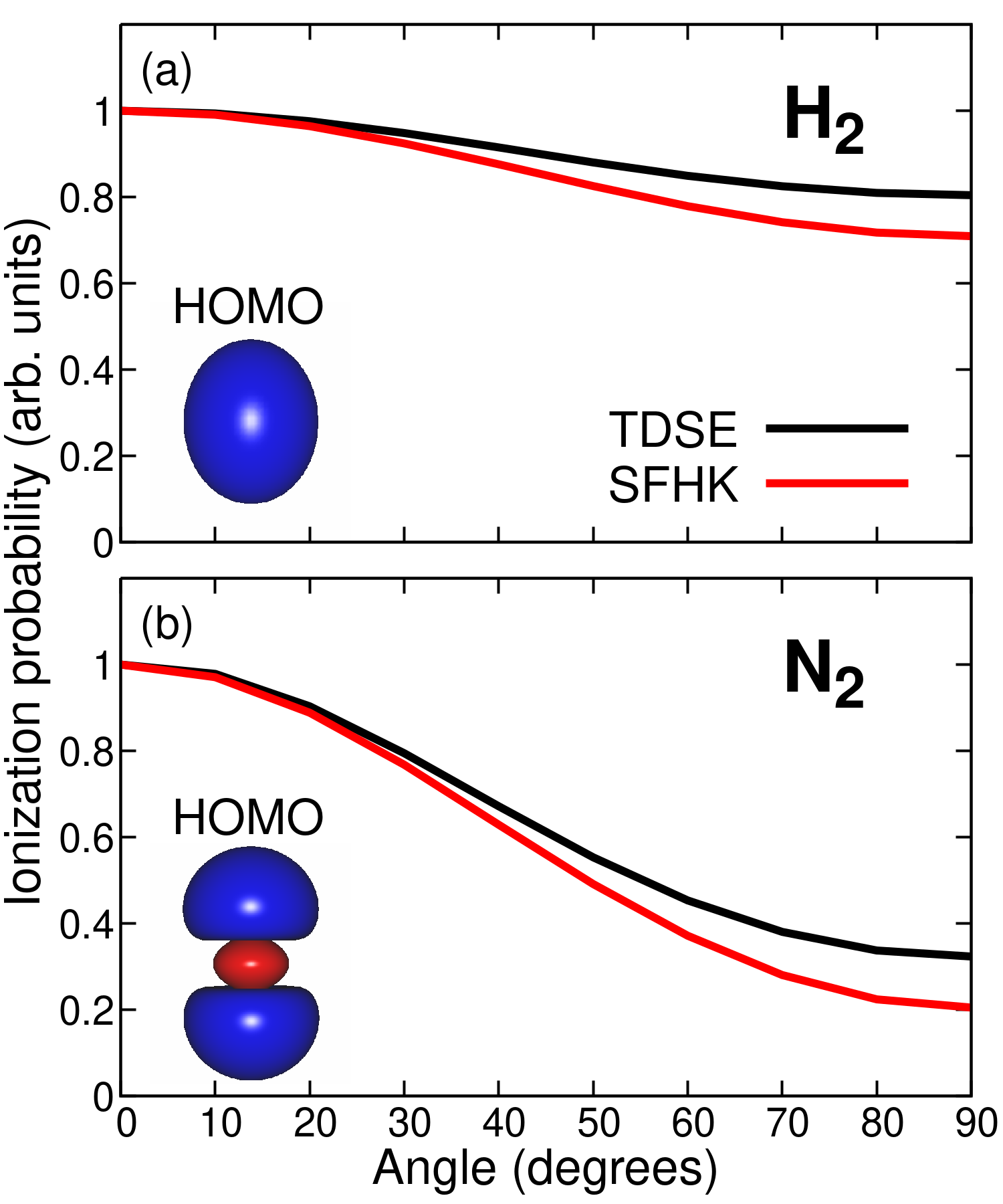}
		\caption{(a): Comparison of ionization yield vs alignment angle from the SFHK (red curve) and the TDSE (black curve) for H$_2$. The yields are normalized at $\theta=0^{\circ}$. (b) The same as in (a), but for N$_2$. The laser parameters are the same as in Fig.~\ref{fig:H2-HHG}.}
		\label{fig:ion-vs-angle}
	\end{center}
\end{figure}

As the SFHK is based on the SFA for the ionization step, it is of interest to compare the angle-dependent ionization probability from the SFA with the TDSE calculation. This comparison is shown in Fig.~\ref{fig:ion-vs-angle} (a) and (b) for H$_2$ and N$_2$, respectively. For both cases, the yields are normalized at $\theta=0^{\circ}$, and the laser parameters are the same as in Fig.~\ref{fig:H2-HHG}. Overall, the agreement between the two methods is quite good, with the largest discrepancy of about $40\%$ near $90^{\circ}$ for N$_2$. Note that the small discrepancy for ionization near $90^{\circ}$ for N$_2$ does not lead to noticeable inaccuracies for HHG yield or phase from the SFHK, as compared with the TDSE, see Fig.~\ref{fig:N2-HHG-angular}. In this regard, we remark that majority of electron that tunneled to the continuum does not return close to the target ion to emit HHG photons. Therefore, a good (or not very good) agreement between the SFA and the TDSE for the {\em total} ionization probability does {\em not} necessarily guarantee the good (or not very good) agreement for the HHG yield and dipole phase.

\subsection{HHG from isotropically distributed H$_2$ and N$_2$}

\begin{figure}[tb!]
	\begin{center}
		\includegraphics[width=0.95\linewidth]{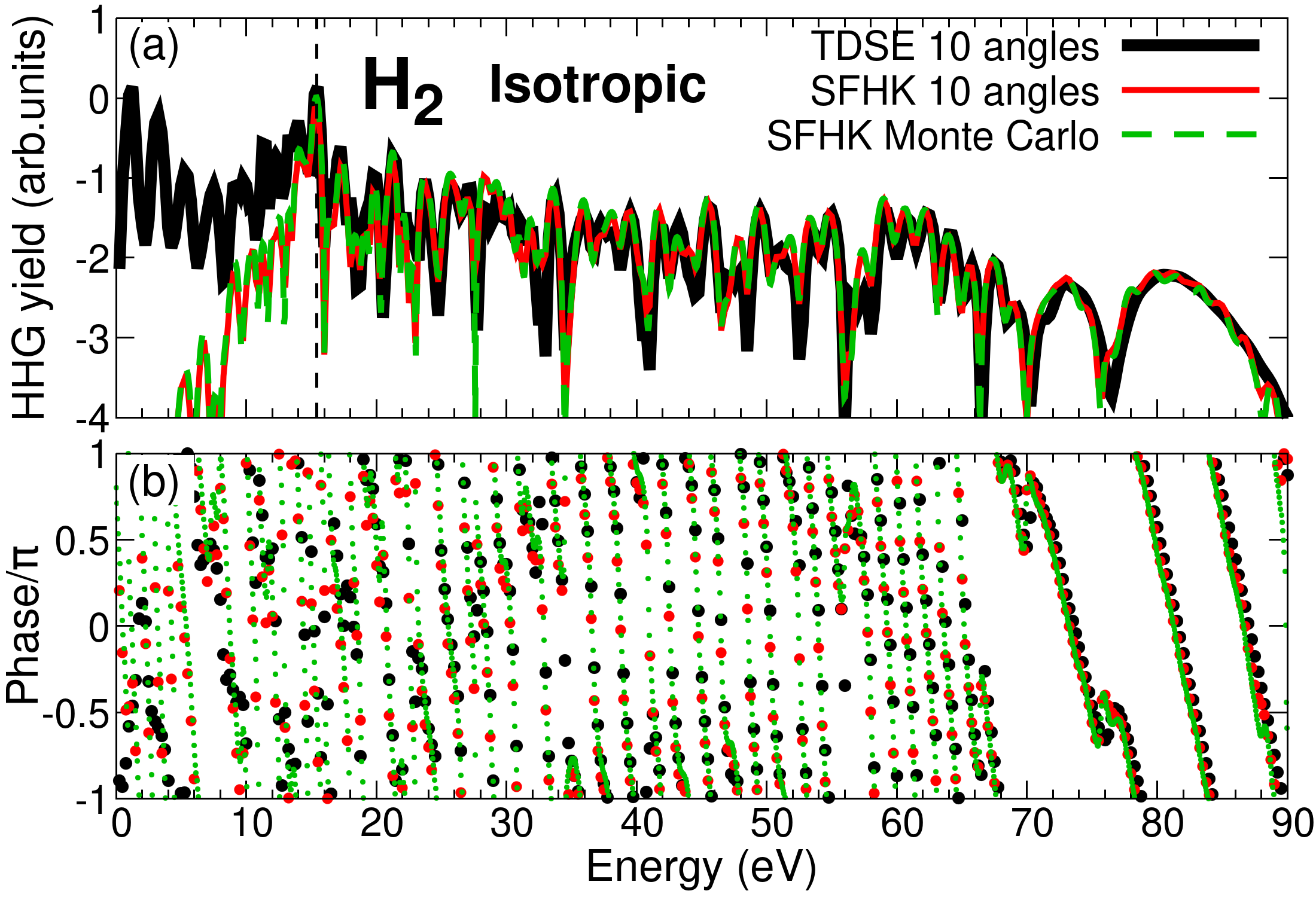}
		\caption{(a) Comparison of HHG spectra from the SFHK (red curve) and the TDSE (black curve) for isotropically distributed H$_2$. The SFHK result obtained with the Monte-Carlo integration over the Euler angles is also shown (green dashed curve). The vertical black dashed line shows the position of the threshold near 15.4 eV. The HHG yields are shown in the logarithm scale. (b) The same as (a), but for the induced dipole phase. The laser parameters are the same as in Fig.~\ref{fig:H2-HHG}.}
		\label{fig:H2-HHG-avg}
	\end{center}
\end{figure}

Once HHG induced dipole is obtained for fixed alignment angles, either from the SFHK or the TDSE, the average over the alignment distribution can be performed in a straightforward manner. In the following, we assume that the molecules are not aligned. That is, they are assumed to be isotropically distributed. Due to the cylindrical symmetry, we found that quite converged alignment averaging can be carried out with as little as ten alignment angles between the molecular axis and laser polarization direction. For the isotropically distributed H$_2$, a comparison between the SFHK and the TDSE is shown in Fig.~\ref{fig:H2-HHG-avg} for H$_2$, indicating excellent agreement between the two methods. This is not surprising, as good agreements were seen for all the fixed alignments, as discussed earlier in Sec.III.1. Within the SFHK, since each alignment angle requires about $10^8$ trajectories, we need about $10^9$ trajectories to get a converged spectrum for isotropically distributed H$_2$.   

More importantly, as discussed in Sec.~II.2, within the SFHK method one can combine the averaging over the alignment distribution (i.e., the integration over the Euler angles specifying the molecular orientation in the laboratory frame) with the standard SFHK integration over the initial momentum distribution of the continuum electron at the tunnel exit. The result of that calculation is also shown in Fig.~\ref{fig:H2-HHG-avg}, which includes about $2 \times 10^8$  trajectories. Note that due to the nature of the Monte-Carlo sampling, each of these trajectories, in principle, has a different set of Euler angles $(\alpha,\beta,\gamma)$. For the practical implementation, this is a factor of 5 more economical than the standard method (which performs the averaging after the induced dipole from the SFHK is obtained, say, for ten different alignment angles). Clearly, all the three methods agree well, for both HHG yield and phase.  

\begin{figure}[tb!]
	\begin{center}
		\includegraphics[width=0.95\linewidth]{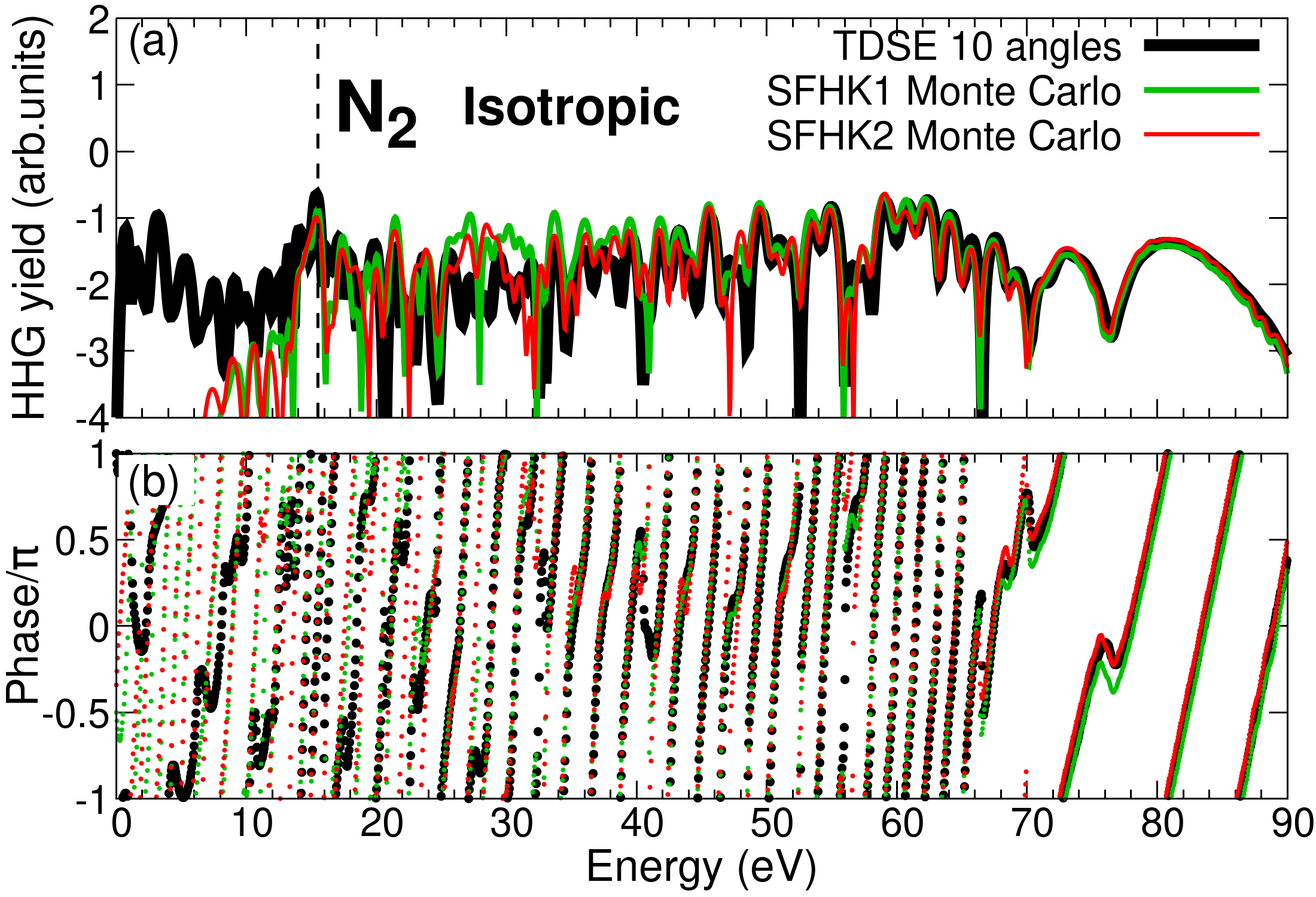}
		\caption{(a) Comparison of HHG spectra from the standard SFHK (denoted as SFHK1 in the label, green curve) and the TDSE (black curve) for isotropically distributed N$_2$. The SFHK with a modified SFA treatment for the ionization step is also shown (denoted as SFHK2 in the label, red curve). The vertical black dashed line shows the position of the threshold near 15.5 eV. The HHG yields are shown in the logarithm scale. (b) The same as (a), but for the induced dipole phase. The laser parameters are the same as in Fig.~\ref{fig:H2-HHG}.}
		\label{fig:N2-HHG-avg}
	\end{center}
\end{figure}

Similar good agreement between the SFHK and the TDSE was found for isotropically distributed N$_2$ case, see Fig.~\ref{fig:N2-HHG-avg}. Here, only the SFHK result obtained with the Monte-Carlo integration over the Euler angles is shown, while for the TDSE, only ten alignment angles were used. Note that there are some discrepancies in the energy range of $[25:40]$ eV. We have checked that the discrepancies are not due to the Monte-Carlo integration method, as the standard method using ten alignment angles gave nearly the same result (not shown). In fact, they are the consequences of the inaccuracies of the SFHK for small angle near $\theta=0^{\circ}$ in this energy range.   

For completeness, we also show in Fig.~\ref{fig:N2-HHG-avg} the SFHK result using the modified SFA treatment for the ionization step, denoted as SFHK2 in the figure label. This version of the SFHK will be discussed in more detail in the next subsection. Here, we simply note that SFHK2 result indeed agrees better with the TDSE in the energy range of $[25:40]$ eV, while the two SFHK versions are nearly the same outside that energy range. Note that both versions of the SFHK shown in the figure were obtained by using the Monte-Carlo integration over the Euler angles. For each version, we used about $2 \times 10^8$ trajectories to get the converged result. This is a factor of 5 less trajectories, as compared to the more standard integration method which employs Eq.~(\ref{vz_int}) directly. In other words, averaging over the alignment distribution for homonuclear diatomic molecules is only about two times more expensive than for a fixed alignment. This is consistent with the H$_2$ cases discussed above.        

We remark that the improvement by a factor of 5 in terms of the number of trajectories reported here for the case of isotropically distributed H$_2$ and N$_2$ using the method presented in Sec.~III.2 might seem like a small improvement. However, the true power of the Monte-Carlo integration method will be more evident for the case of polyatomic molecules. In fact, we have found that one typically needs about $2 \times 10^8$ trajectories to get a converged HHG spectrum for a typical polyatomic molecular target with a fixed orientation, and about $5 \times 10^8$ trajectories for the same target with isotropic distribution \cite{benzene:2026}. In other words, averaging over the alignment distribution for polyatomic case is only about three times more expensive than for a fixed alignment. 

\subsection{The nature of the discrepancies for N$_2$ at small alignment angles}

\begin{figure}[tb!]
	\begin{center}
		\includegraphics[width=0.95\linewidth]{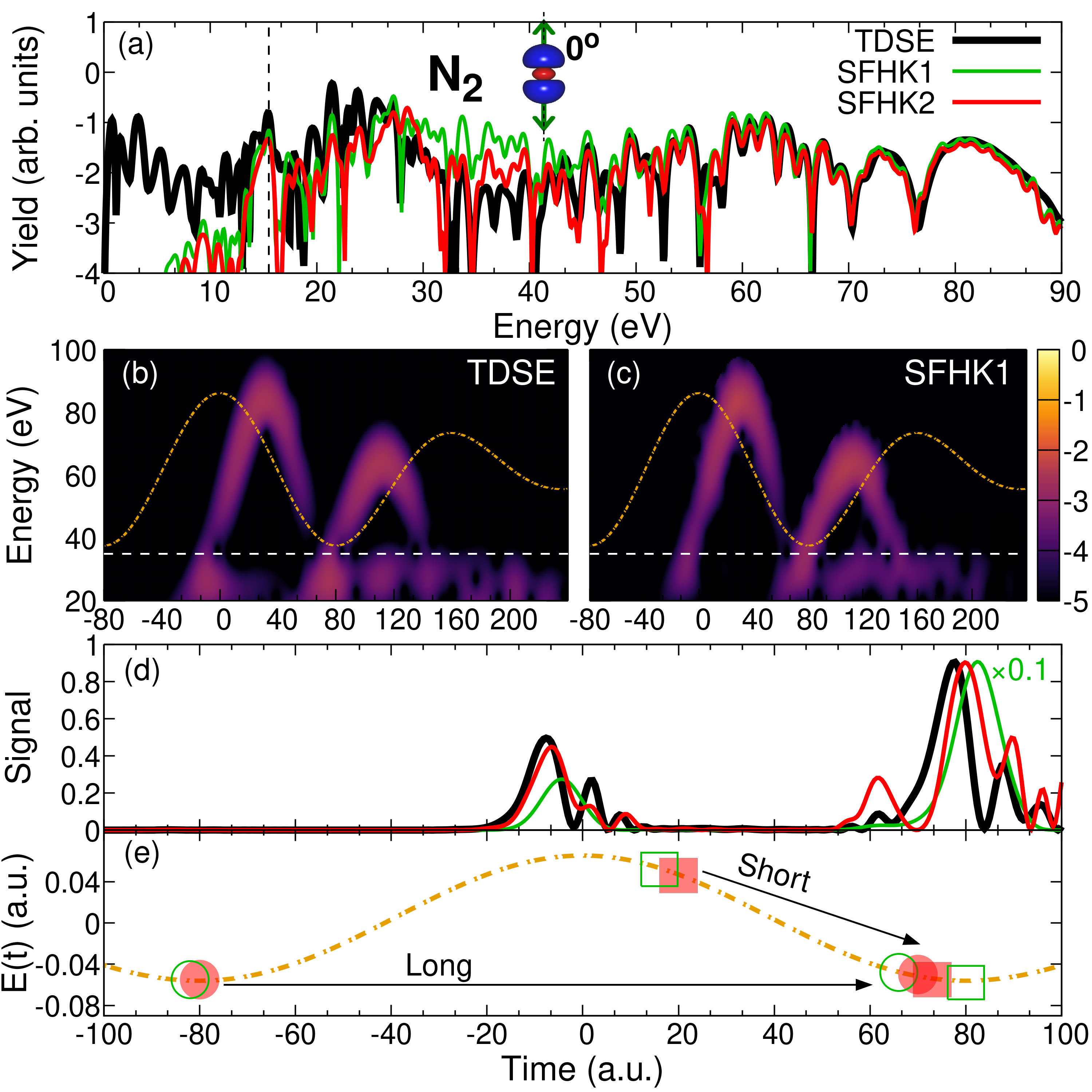}
		\caption{(a) Comparison of HHG spectra from the SFHK (denoted as SFHK1 in the label, red curve) and the TDSE (black curve) for N$_2$ at the alignment angle $\theta=0^{\circ}$. Result from a modified version of the SFHK (i.e., SFHK2) is also shown. The vertical black dashed line shows the position of the threshold near 15.5 eV. The HHG yields are shown in the logarithm scale.  (b) and (c) The Gabor time-frequency analysis of the induced dipole velocity from the TDSE and SFHK, respectively. (d) A slice of the Gabor transform at the energy of $35$ eV. This slice is indicated by the horizontal dashed line in panel (b) and (c). (e) A schematic of long and short trajectories. The long (or short) trajectory is given as circle (square). The trajectories from the modified (or standard) SFA are given as filled (empty) symbols. Dashed-dotted line in (b), (c), and (e) shows the laser electric field. The laser parameters are the same as in Fig.~\ref{fig:H2-HHG}.}
		\label{fig:N2-theta0}
	\end{center}
\end{figure}

In this subsection, we will analyze in more detail the HHG from N$_2$ at $\theta=0$. First, we note that the HHG spectrum from the SFHK agrees well with the TDSE, except in the energy range of $[25:40]$ eV, see Fig.~\ref{fig:N2-theta0}(a). The nature of this discrepancy is not fully understood currently. We remark that the SFHK result is already quite stable with respect to the number of trajectories and other parameters. Note that within the current implementation of the SFHK, we neglect the depletion effect as well as the Stark shift of the ground state. This assumption can be checked using the TDSE. In fact, within the TDSE, instead of calculating the exact induced dipole as $d(t)=\left<\Psi(t)|z|\Psi(t)\right>$, we can use an approximate $\tilde{d}(t)=\left<\Psi_0(t)|z|\Psi(t)\right> + c.c.$ Here, $\left|\Psi(t)\right>$ is the state vector that evolves in time in the laser field from the initial state $\left|\Psi_0\right>$. Clearly, the depletion and Stark shift are neglected in $\left|\Psi_0(t)\right>=\left|\Psi_0\exp(-iE_0t)\right >$. We found that HHG spectra from the TDSE using the approximate $\tilde{d}(t)$ is nearly identical to that from the exact $d(t)$ (not shown).This indicates that the SFHK assumption is adequate and it is not the source of the discrepancies.
A similar version using dipole velocity was also used.  

\begin{figure}[tb!]
	\begin{center}
		\includegraphics[width=0.95\linewidth]{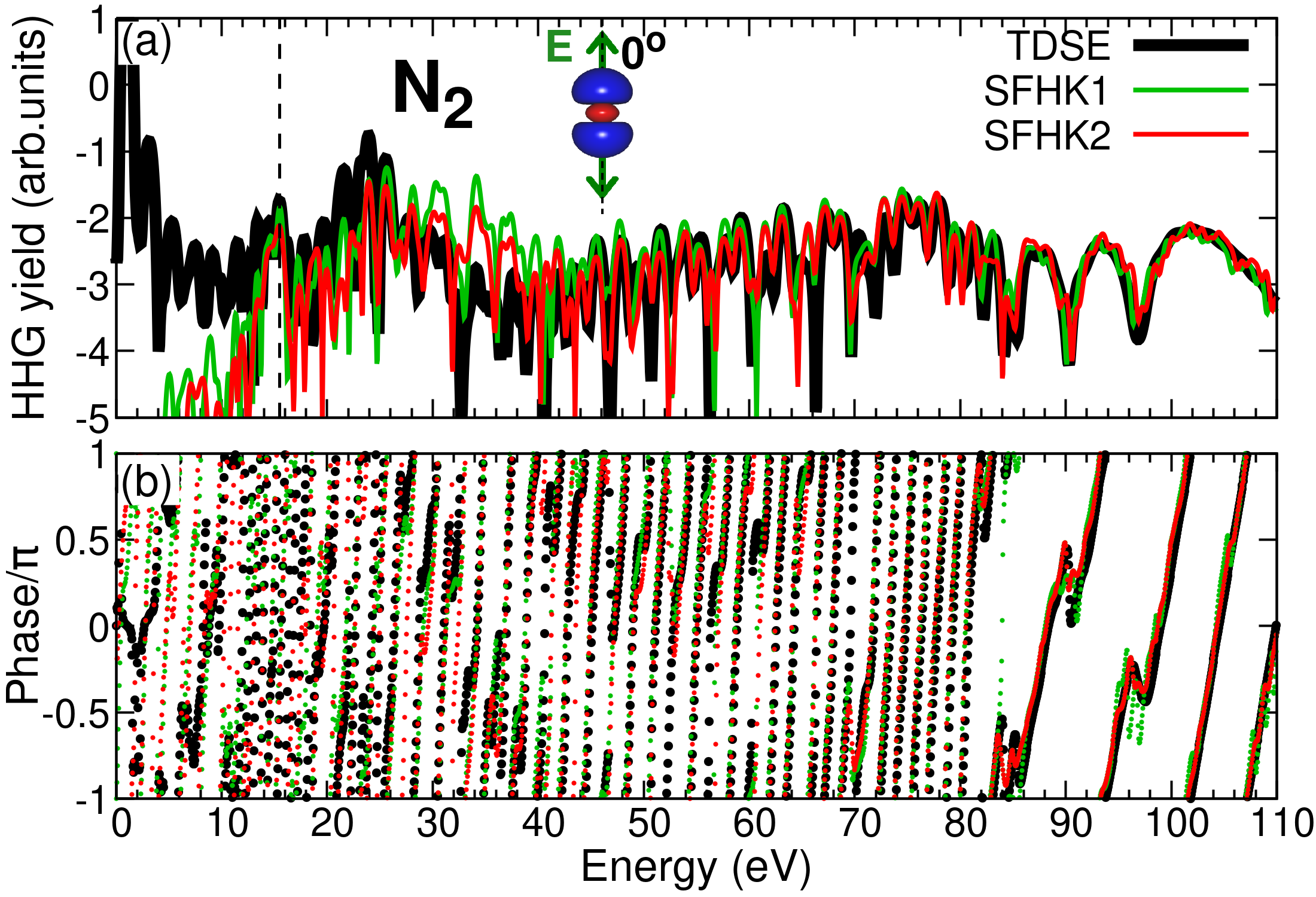}
		\caption{(a) Same as Fig.~\ref{fig:N2-theta0}(a), but for laser intensity of $2\times 10^{14|}$ W/cm$^2$. The induced dipole phase is also shown in (b).}
		\label{fig:N2-theta0-2I0}
	\end{center}
\end{figure}

We further found that the above discrepancies in the same energy range occur also for other laser parameters. As an example, we show in Fig.~\ref{fig:N2-theta0-2I0} for the laser intensity of $2\times 10^{14}$ W/cm$^2$, while other laser parameters are kept the same as in Fig.~\ref{fig:N2-theta0}(a).

To understand the origin of these discrepancies, we compare in Fig.~\ref{fig:N2-theta0}(b) and (c) the Gabor  time-frequency analysis of the induced dipole velocity from the TDSE and SFHK, respectively. Overall, the two results are quite similar. The largest difference is indeed in the energy range of [25:40] eV within the time window $t=[60:90]$ a.u., where the long and short trajectories interfere. To see that more clearly, we show in Fig.~\ref{fig:N2-theta0}(d) a slice of the Gabor transform at $E=35$ eV. Here, the SFHK result has been rescaled by a factor of 0.1. The SFHK result is also slightly delayed as compared to the TDSE. The long trajectories are emitted to the continuum near $t=-80$ a.u. They interfere with the short trajectories that are ``born'' into the continuum near $t=20$ a.u. (i.e., in the next sub-cycle after the long trajectories were born) -- see Fig.~\ref{fig:N2-theta0}(e). These trajectories returns to the target ion in the time window $t=[60:90]$ a.u.    

We speculate that the possible reason for the discrepancies is due to the inaccuracies of the SFA treatment for the ionization step within the saddle-point approximation. We therefore propose that the saddle-point equation for ionization Eq.~(\ref{SP-eq}) to be modified as
\begin{equation}
	\frac{[{\bf{k}} + {\bf{A}}(t_s) -  {\bf{A}}(t_0)]^2}{2}+I_p -\eta{\bf{E}}(t_s)\cdot{\bf{r}}_0= 0.
	\label{MSP-eq}
\end{equation}
with $\eta=1$ (or $\eta=-1$) for the electron emitted from the left (right) side, and ${\bf{r}}_0={\bf{R}}/2$. Note that this modification has been suggested earlier in Ref.~\cite{Chirila:pra06} and was also used recently in Ref.~\cite{Xie:prl21}. The version of the SFHK with the modified SFA of Eq.~(\ref{MSP-eq}) will be called SFHK2 in the following. Since $r_0 \approx 1$ is quite small, the influence of the extra term $-\eta{\bf{E}}(t_s)\cdot{\bf{r}}_0$ is to slightly increase the ``effective'' ionization potential. It therefore slightly changes the born time and tunnel exit position, as compared to the standard SFA treatment. The modification to the trajectories are shown schematically in Fig.~\ref{fig:N2-theta0}(e). Note that the effect is strongest for $\theta=0$ and vanishes at $\theta=90^{\circ}$ where ${\bf{E}}\perp {\bf{r}}_0$. The results of the SFHK2 are shown in Fig.~\ref{fig:N2-theta0}(a). Overall, we found that the SFHK2 agree better with the TDSE. This is in agreement with the recent finding by Xie {\it et al} \cite{Xie:prl21}. Fig.~\ref{fig:N2-theta0}(d) shows more detail of the improvement for the slice of the Gabor transform at $35$ eV. In fact, the main effect is the significant suppression of the returning electron near $t=[60:90]$ a.u. Furthermore, the return time window of the long and short trajectories in the SFHK2 have a broader overlap, leading to a stronger interference oscillation of the Gabor transform, which resemble more closely the TDSE result. 

Similar improvement was also found for the SFHK2 result for the higher laser intensity case, see Fig.~\ref{fig:N2-theta0-2I0}, and somewhat weaker improvement was found for the case of isotropic distribution, see Fig.~\ref{fig:N2-HHG-avg}. For the H$_2$ case, the SFHK2 does not lead to significant changes.

\section{Summary}

In conclusion, we have extended the strong-field Herman-Kluk propagator (SFHK) method for calculations of HHG spectra and induced harmonic phases to molecular targets. We have demonstrated on the example of H$_2$ and N$_2$ very high accuracy of the SFHK for typical laser parameters in the tunneling regime. Due to the semiclassical trajectory-based nature, the SFHK can provide intuitive pictures of the HHG process. From a practical standpoint, the SFHK is also very appealing as primitive parallelization can be easily implemented for independent trajectories. In this paper, we have shown that ensemble averaging over the molecular alignment can be combined efficiently together with the integral over the initial momentum distribution. This effectively reduces the number of trajectories needed for practical calculations. We expect that this advantage of the SFHK is especially important for polyatomic targets.   

We have also found that the SFHK can also be used to provide a severe test for the accuracy of the SFA treatment of the strong-field ionization. In fact, the use of the standard saddle-point approximation of the SFA for the ionization step within the SFHK lead to somewhat inaccurate HHG spectrum in N$_2$ at small alignment angles. But the same approximation work well for H$_2$ and all other atoms that we tried \cite{Tran:HHGatom-2026}. While the result of SFHK2, which employs a modified version of the SFA, has improve the results in N$_2$, further investigation is strongly desired. This is of critical importance, as the tunneling ionization is at heart of many strong-field phenomena such as HHG, high-order above threshold ionization and laser-induced electron diffraction, as well as nonsequential double ionization.   

In the future, the SFHK can be extended to include the nuclear motion and to treat coupled electron-nuclear dynamics in strong-field physics phenomena, in which accurate description of dynamics of electron in the continuum is critical.

\section*{Acknowledgments}
This work was supported by the U.S. Department of Energy (DOE), Office of Science, Basic Energy Sciences (BES) under Award Number DE-SC0023192. We thank Richard Jones and the Storrs HPC facilities for valuable computational supports and resources. This research was done using services provided by the OSG Consortium \cite{osg07new,osg09new,osg06new,osg05new}.

%\bibliography{MyBib,Phi_Bib} 

\end{document}